\documentclass[11pt]{article}
\usepackage{amssymb}
\usepackage{cite}
\usepackage{graphicx}
\usepackage{amsmath}
\usepackage{indentfirst}
\usepackage{subfigure}
\begin{document}
\parindent 14pt
\begin{center}
\baselineskip=14pt {\Large\bf Approximate homotopy symmetry method
and homotopy series solutions to the six-order boussinesq equation}
\end{center}
 \begin{description}
\item[{}\mbox{}\quad]
\begin{center}
Xiaoyu Jiao$^{a,}$\footnote{Corresponding Author: Xiaoyu Jiao,
jiaoxyxy@yahoo.com.cn}, Yuan Gao$^{b}$ and S. Y. Lou$^{a,b,c}$
\end{center}
\begin{center}
{\small $^a$Department of Physics, Shanghai Jiao Tong
University, Shanghai, 200240,  China\\
$^b$School of Mathematics, Fudan University, Shanghai, 200433, China\\
$^c$Department of Physics, Ningbo University, Ningbo, 315211, China
}
\end{center}
{\bf Abstract:} An approximate homotopy symmetry method for
nonlinear problems is proposed and applied to the six-order
boussinesq equation which arises from fluid dynamics. We summarize
the general formulas for similarity reduction solutions and
similarity reduction equations of different orders, educing the
related homotopy series solutions. Zero-order similarity reduction
equations are equivalent to Painlev\'e IV type equation or
Weierstrass elliptic equation. Higher order similarity solutions can
be obtained by solving linear variable coefficients ordinary
differential equations. The convergence region of homotopy series
solutions can be adjusted by the auxiliary parameter. Series
solutions and similarity reduction equations from approximate
symmetry method can be retrieved from approximate homotopy symmetry
method.
 \end{description}

\section{Introduction}
Nonlinear phenomena arise in many aspects of science and
engineering. Lie group theory \cite{Lie,Olver,Bluman1,Bluman2}
provides remarkable techniques in effectively studying nonlinear
problems such as exploring similarity reduction of partial
differential equations. It should be noted that approximate
solutions also contribute to understanding the essence of
nonlinearity. Perturbation theory \cite{Cole,Dyke,Nayfeh} was
consequently developed and it plays an essential role in nonlinear
science, especially in finding approximate analytical solutions to
perturbed partial differential equations.

Combined with Lie group theory, perturbation theory gives rise to
two distinct approximate symmetry methods. For the first method due
to Baikov, et al \cite{Baikov1,Baikov2}, symmetry group generators
are generalized to perturbation forms. The second method proposed by
Fushchich, et al \cite{Fushchich} is based on the perturbation
series for the dependent variables which decomposes the original
equation into a system of equations. Approximate symmetry of the
original equation boils down to exact symmetry of such system of
equations resulted from perturbation. The comparisons in Refs.
\cite{Pakdemirli,Wiltshire} show superiority of the second method to
the first one.

Aside from perturbation theory, some nonperturbative techniques,
such as the artificial small parameter method \cite{Lyapunov}, the
$\delta$-expansion method \cite{Karmishin} and the Adomian's
decomposition method \cite{Adomian} are also of significance when
perturbation quantities are not involved in many problems.

Liao \cite{Liao} developed a kind of analytic technique, the
homotopy analysis method, and solved some problems successfully
\cite{Liao1,Liao2,Liao3}. Recently, the homotopy analysis method was
further improved in Refs. \cite{Liao4,Liao5,Liao6}. The homotopy
analysis method is based on homotopy conception in topology. This
method is suitable for problems that contain no small parameters.
Furthermore, the series solutions obtained from the perturbation
method, the artificial small parameter method, the
$\delta$-expansion method and the Adomian's decomposition method can
also be retrieved by the homotopy analysis method.

Perturbation techniques are only confined to weak perturbation
problems. For strongly perturbed nonlinear systems, we propose
approximate homotopy symmetry method to study possible analytic
series solutions. The six-order boussinesq equation is used as an
example to illustrate the effectiveness of the homotopy symmetry
method.
\section{Basic notions}
For a nonlinear partial differential equation
\begin{equation}\label{original}
\mathcal {A}(u)=\mathcal {A}(x,t,u_x,u_t,u_{xx},u_{xt},\cdots)=0,
\end{equation}
where $\mathcal {A}$ is a nonlinear operator, $u=u(x,t)$ is an
undetermined function, and $\{x,\ t\}$ are independent variables, we
introduce a homotopy model
\begin{equation}\label{homotopy}
\mathcal {H}(u,q)=0,
\end{equation}
with $q\in[0,\ 1]$ an embedding homotopy parameter. The above
homotopy model has the property
\begin{equation}\label{homotopy01}
\mathcal {H}(u,1)=\mathcal {A}(u),\ \mathcal {H}(u,0)=\mathcal
{H}_0(u),
\end{equation}
where $\mathcal {H}_0(u)$ is a differential equation of which the
solutions can be easily obtained.

We make an ansatz that the homotopy model \eqref{homotopy} has the
homotopy series solution
\begin{equation}\label{series}
u=\sum_{i=0}^\infty u_iq^i,
\end{equation}
where $u_i$ solves the system
\begin{eqnarray}
&&  \mathcal {H}_0(u_0)=0,\label{homotopys0}\\
&&  \mathcal {H}'_0(u_0)u_1+F_1(u_0)=0,\label{homotopys1}\\
&&  \mathcal {H}'_0(u_0)u_2+F_2(u_0,\ u_1)=0,\\
&& \nonumber \cdots \cdots, \\
&&  \mathcal {H}'_0(u_0)u_i+F_i(u_0,\ u_1,\ \cdots,\ u_{i-1})=0,\label{homotopysi}\\
&& \nonumber \cdots \cdots,
\end{eqnarray}
in which the operator $\mathcal {H}'_0(u_0)$ is defined as
$$\mathcal {H}'_0(u_0)f=\left.\frac{\partial} {\partial \varepsilon}\mathcal
{H}_0\left(u_0+\varepsilon f\right)\right|_{\varepsilon=0} $$ for
arbitrary function $f(x,t)$, and all
 $F_i\equiv F_i(u_0,\ u_1,\ \cdots,\ u_{i-1})$ satisfy
\begin{eqnarray}
F_i=\left.\frac1{i!}\frac{\partial^i} {\partial q^i}\mathcal
{H}\left(\sum_{k\neq i} u_kq^k,q\right)\right|_{q=0},\ (i=1,\ 2,\
\cdots).
\end{eqnarray}
Then, the solutions of the original nonlinear system
\eqref{original} read
\begin{equation}\label{series1}
u=\sum_{i=0}^\infty u_i.
\end{equation}

Now, we are concerned about constructing approximate group invariant
solutions of the homotopy model \eqref{homotopy}.
First, we introduce the following definitions:\\
\bf Definition 1. \rm \em Symmetry (or exact symmetry). \rm A
symmetry, $\sigma$, of the nonlinear equation \eqref{original} is
defined as a solution of its linearized equation
\begin{equation}\label{Es}
\mathcal {A}'(u)\sigma\equiv \left.\frac{\partial}{\partial
\varepsilon}\mathcal {A}[u+\varepsilon
\sigma]\right|_{\varepsilon=0}=0,
\end{equation}
which means that Eq. \eqref{original} is form invariant under the
transformation $u\rightarrow u+\varepsilon \sigma$ with an
infinitesimal parameter $\varepsilon$.
\\
\bf Definition 2. \rm \em Homotopy symmetry. \rm A homotopy
symmetry, $\sigma_q$, of the nonlinear system \eqref{original} is an
exact symmetry of the related homotopy model \eqref{homotopy}, i.e.,
a solution to the linearized equation of \eqref{homotopy}
\begin{equation}\label{Ehs}
\mathcal {H}'(u,q)\sigma_q=0.
\end{equation}
\bf Definition 3. \rm \em Approximate homotopy symmetry. \rm The
$k$th order approximate homotopy symmetry,
$\sigma_{q,k}=\left(\sigma_{0},\ \sigma_{1},\ \sigma_{2},\cdots,\
\sigma_k\right), $ of the nonlinear system \eqref{original} is an
exact symmetry of the system of the first $(k+1)$ approximate
equations, i.e., the solution of the following linearized system
\begin{eqnarray}
&&\mathcal {H}_0'(u_0)\sigma_0=0,\label{E0}\\
&&\mathcal {H}_0'(u_0)\sigma_k+\left(\mathcal
{H}_0'(u_0)\right)'\sigma_0u_k+\sum_{j=0}^{k-1}\left(F_{k}\right)'_{u_j}\sigma_j=0,\
(j=1,\ 2,\ ...,\ k),
\end{eqnarray}
where the operator $\left(F_{k}\right)'_{u_j}$ is defined as
$$\left(F_{k}\right)'_{u_j}=\left.\frac{\partial}{\partial \varepsilon}{F_k}(u_0,\ u_1,\ ...,\ u_j+\varepsilon
\sigma_j,\ u_{j+1},\ ...,\ u_{k-1})\right|_{\varepsilon=0}.$$

For simplicity later, the following simple homotopy model is
exclusively taken
\begin{equation}\label{homotopyeq}
(1-q)\mathcal {H}_0(u)+q\lambda\mathcal {A}(u)=0,
\end{equation}
with $\lambda\neq0$ an auxiliary parameter. It is easily seen that
Eq. \eqref{homotopyeq} varies asymptotically from $\mathcal
{H}_0(u)=0$ to Eq. \eqref{original} as $q$ goes gradually from 0 to
1. When $\mathcal {H}_0$ is fixed as a linear operator, the homotopy
model \eqref{homotopy} is just the usual one applied in Refs.
\cite{Liao,Liao1,Liao2,Liao3,Liao4,Liao5,Liao6}.

From the above process, we see that the approximate homotopy
symmetry method is an integration of the homotopy concept,
perturbation analysis and symmetry method.
\section{Approximate homotopy symmetry method to the six-order boussinesq equation}
The illposed Boussinesq equation
\begin{equation}\label{bouss1}
u_{tt}=(u+u^2)_{xx}+u_{xxxx},
\end{equation}
describes propagation of long waves in shallow water under gravity
\cite{Whitham}, in one-dimensional nonlinear lattices and in
nonlinear strings \cite{Zakharov}. Filtering and regularization
techniques was applied to Eq. \eqref{bouss1} in Ref. \cite{Daripa1}
to introduce the singularly perturbed (sixth-order) Boussinesq
equation
\begin{equation}\label{bouss2}
\eta_{tt}=\eta_{xx}+(\eta^2)_{xx}+\eta_{xxxx}+\epsilon^2\eta_{xxxxxx},
\end{equation}
where $\epsilon$ is a small parameter. In Ref. \cite{Daripa2},
double-series perturbation analysis was utilized to recover Eq.
\eqref{bouss2}.

For formal brevity, we rewrite Eq. \eqref{bouss2} as
\begin{equation}
u_{tt}+u_{xx}+(u^2)_{xx}+u_{xxxx}=\mu u_{xxxxxx},\label{bouss3}
\end{equation}
where $u$ is a function with respect to $x$ and $t$, $\mu$ is an
arbitrary parameter. Assuming $\mathcal {H}_0(u)$ to be the left
hand side of Eq. \eqref{bouss3} and replacing $\lambda$ by
$1-\theta$ for concision of the results, we change Eq.
\eqref{homotopyeq} into
\begin{equation}\label{homotopyeq1}
(1-q\theta)(u_{tt}+u_{xx}+(u^2)_{xx}+u_{xxxx})-q\mu(1-\theta)u_{xxxxxx}=0.
\end{equation}
It is easily seen that Eq. \eqref{homotopyeq1} is just the
boussinesq equation \cite{Boussinesq1,Boussinesq2} when $q=0$.

Substituting Eq. \eqref{series} into the above equation and matching
the coefficients of different powers of $q$ yield the following
system of partial differential equations ($k$-order approximate
equation)
\begin{equation}\label{deformationeq}
u_{k,tt}+u_{k,xx}+2\sum_{i=0}^k(u_{k-i,x}u_{i,x}+u_{k-i}u_{i,xx})+u_{k,xxxx}+
\mu(\theta-1)\sum_{i=0}^{k-1}\theta^{k-1-i}u_{i,xxxxxx}=0,\ (k=0,\
1,\ \cdots)
\end{equation}
with $u_{-1}=0$.

We investigate similarity reduction of Eq. \eqref{deformationeq} by
Lie symmetry method \cite{Lou}. The linearized equations for Eq.
\eqref{deformationeq} are
\begin{eqnarray}
&&\sigma_{k,tt}+\sigma_{k,xx}+2\sum_{i=0}^k(\sigma_{k-i,x}u_{i,x}+u_{k-i,x}\sigma_{i,x}+
\sigma_{k-i}u_{i,xx}+u_{k-i}\sigma_{i,xx})\nonumber\\
&&+\sigma_{k,xxxx}+\mu(\theta-1)\sum_{i=0}^{k-1}\theta^{k-1-i}\sigma_{i,xxxxxx}=0,\
(k=0,\ 1,\ \cdots)\label{linearizedeq}
\end{eqnarray}
where $\sigma_k$ are functions of $x$ and $t$ with $\sigma_{-1}=0$.
Eq. \eqref{linearizedeq} means that Eq. \eqref{deformationeq} is
invariant under the transformation $u_k\rightarrow
u_k+\varepsilon\sigma_k,\ (k=0,\ 1,\ \cdots)$ with $\varepsilon$ an
infinitesimal parameter.

The symmetry transformations
\begin{equation}\label{sym}
\sigma_k=Xu_{k,x}+Tu_{k,t}-U_k,\ (k=0,\ 1,\ \cdots),
\end{equation}
where $X$, $T$ and $U_k$ are functions with respect to $x$, $t$ and
$u_k,\ (k=0,\ 1,\ \cdots)$, conform to Eq. \eqref{linearizedeq}
under Eq. \eqref{deformationeq}. We consider finite
equations in Eqs. \eqref{deformationeq}, \eqref{linearizedeq} and
\eqref{sym} to summarize general formulas for similarity reduction
solutions and similarity reduction equations.

Confining the maximum of $k$ to 2 in Eqs.
\eqref{deformationeq}, \eqref{linearizedeq} and \eqref{sym}, we see
that the independent variables of $X$, $T$, $U_0$, $U_1$ and $U_2$
are accordingly restricted to $x$, $t$, $u_0$, $u_1$ and $u_2$. More
than 2000 determining equations are obtained by inserting Eq.
\eqref{sym} into Eq. \eqref{linearizedeq}, eliminating $u_{0,tt}$,
$u_{1,tt}$ and $u_{2,tt}$ in terms of Eq. \eqref{deformationeq} and
vanishing coefficients of different partial derivatives of $u_0$,
$u_1$ and $u_2$.

To solve the determining equations, we first extract the simplest
equations for $T$
\begin{displaymath}
T_x=T_{u_0}=T_{u_1}=T_{u_2}=0,
\end{displaymath}
from which we have $T=T(t)$. Considering this condition, we select
the simplest equations for $X$
\begin{displaymath}
X_t=X_{u_0}=X_{u_1}=X_{u_2}=0,
\end{displaymath}
with the solution $X=X(x)$. In this case, we get the simplest
equations for $U_0$, $U_1$ and $U_2$
\begin{displaymath}
U_{0,u_0u_0}=U_{0,u_1}=U_{0,u_2}=0,
\end{displaymath}
\begin{displaymath}
U_{1,tu_0}=U_{1,tu_2}=U_{1,u_0u_0}=U_{1,u_0u_1}=U_{1,u_0u_2}=U_{1,u_1u_1}=U_{1,u_1u_2}=U_{1,u_2u_2}=0,
\end{displaymath}
\begin{displaymath}
U_{2,tu_0}=U_{2,tu_1}=U_{2,u_0u_0}=U_{2,u_0u_1}=U_{2,u_0u_2}=U_{2,u_1u_1}=U_{2,u_1u_2}=U_{2,u_2u_2}=0,
\end{displaymath}
which imply
\begin{displaymath}
U_0=F_1(x,t)u_0+F_2(x,t),
\end{displaymath}
\begin{displaymath}
U_1=F_3(x)u_0+F_4(x,t)u_1+F_5(x)u_2+F_6(x,t),
\end{displaymath}
\begin{displaymath}
U_2=F_7(x)u_0+F_8(x)u_1+F_9(x,t)u_2+F_{10}(x,t),
\end{displaymath}
where the undetermined functions satisfy
\begin{displaymath}
F_3=F_5=F_6=F_7=F_{10}=F_{1,x}=F_{1,t}=F_{2,x}=F_{2,tt}=F_{4,x}=F_{4,t}=F_{8,x}=F_{9,x}=F_{9,t}
\end{displaymath}
\begin{displaymath}
=X_{xx}=T_{tt}=0,\ T_t=2X_x,\ 2X_x=2T_t+F_1,\ X_x=T_t+F_2,\
2X_x=2T_t+2F_4-F_9,
\end{displaymath}
\begin{displaymath}
6X_x=2T_t+F_4-F_9,\ 6X_x=2T_t+F_1-F_4,\ 6\theta X_x=2\theta
T_t+\theta(F_1-F_9)-F_8.
\end{displaymath}
The solutions to the determining equations are finally obtained by
solving the above system as follows
\begin{equation}
X=\frac{1}{2}C_1x+C_3,\ T=C_1t+C_2,\ U_0=-C_1u_0-\frac{1}{2}C_1,\
U_1=-2C_1u_1,\ U_2=C_1(\theta u_1-3u_2),
\end{equation}
where $C_1$, $C_2$ and $C_3$ are arbitrary constants.

In the same way, limiting the maximum of $k$ to 3 in Eqs. \eqref{deformationeq}, \eqref{linearizedeq} and
\eqref{sym}, we execute similar computation and obtain
\begin{eqnarray}
&&X=\frac{1}{2}C_1x+C_3,\ T=C_1t+C_2,\ U_0=-C_1u_0-\frac{1}{2}C_1,\nonumber\\
&&U_1=-2C_1u_1,\ U_2=C_1(\theta u_1-3u_2),\ U_3=2C_1(\theta u_2-2u_3),
\end{eqnarray}
where $C_1$, $C_2$ and $C_3$ are arbitrary constants.

Enlarge the domain of $k$ by degrees and repeat similar procedures
, we discover the formal coherence of $X$, $T$ and $U_k\ (k=0,\ 1,\ \cdots)$
, i.e.,
\begin{equation}\label{infinitesimalgenerator}
X=\frac{1}{2}C_1x+C_3,\ T=C_1t+C_2,\ U_k=C_1[(k-1)\theta
u_{k-1}-(k+1)u_k]-\frac{1}{2}C_1\delta_{k,0},\ (k=0,\ 1,\ \cdots)
\end{equation}
where $C_k,\ (k=0,\ 1,\ \cdots)$ are arbitrary constants. The notation
$\delta_{k,0}$ satisfying $\delta_{0,0}=1$ and $\delta_{k,0}=0\
(k\neq0)$ is adopted in the following text. Similarity solutions to
Eq. \eqref{deformationeq} can be obtained by solving the
characteristic equations
\begin{equation}\label{charaeq}
\frac{{\rm d}x}{X}=\frac{{\rm d}t}{T},\ \frac{{\rm d}u_0}{U_0}
=\frac{{\rm d}t}{T},\ \frac{{\rm d}u_1}{U_1}=\frac{{\rm d}t}{T},\
\cdots,\ \frac{{\rm d}u_k}{U_k}=\frac{{\rm d}t}{T},\ \cdots
\end{equation}
which are distinguished in two subcases.
\subsection{Homotopy symmetry reduction of Painlev\'e IV type}
When $C_1\neq0$, without loss of generality, we rewrite the
constants $C_2$ and $C_3$ as $C_1C_2$ and $\frac{1}{2}C_1C_3$ and
change Eq. \eqref{infinitesimalgenerator} to
\begin{equation}\label{infinitesimalgenerator1}
X=\frac{1}{2}C_1(x+C_3),\ T=C_1(t+C_2),\ U_k=C_1[(k-1)\theta
u_{k-1}-(k+1)u_k]-\frac{1}{2}C_1\delta_{k,0},\ (k=0,\ 1,\ \cdots).
\end{equation}

From the first three equations in Eq. \eqref{charaeq}, we get the
invariants
\begin{eqnarray}
&&I(x,t)=\xi=\frac{x+C_3}{\sqrt{t+C_2}},\\
&&I_0(x,t,u_0)=P_0=u_0t+C_2u_0+\frac{1}{2}t,\\
&&I_1(x,t,u_1)=P_1=(t+C_2)^2u_1.
\end{eqnarray}
Viewing $P_0$ and $P_1$ as functions of $\xi$, we have
\begin{eqnarray}
&&u_0=\frac{2P_0(\xi)-t}{2(t+C_2)},\\
&&u_1=\frac{P_1(\xi)}{(t+C_2)^2}.
\end{eqnarray}
Similarly, we get other similarity solutions
\begin{eqnarray}
&&u_2=\frac{P_2(\xi)}{(t+C_2)^3}+\frac{\theta
tP_1(\xi)}{(t+C_2)^3},\\
&&u_3=\frac{\theta^2t^2P_1(\xi)}{(t+C_2)^4}+\frac{2\theta
tP_2(\xi)}{(t+C_2)^4}+\frac{P_3(\xi)}{(t+C_2)^4},\\
&&\cdots\ \cdots\ \cdots\nonumber
\end{eqnarray}
which conform to the general expression
\begin{equation}\label{similaritysol1}
u_k=\frac{2P_0(\xi)-t}{2(t+C_2)}\delta_{k,0}+\sum_{i=0}^{k-1}\tbinom{k-1}{i}\frac{(\theta
t)^iP_{k-i}(\xi)}{(t+C_2)^{k+1}},\ (k=0,\ 1,\ \cdots)
\end{equation}
with the similarity variable $\xi=\frac{x+C_3}{\sqrt{t+C_2}}$.

From Eq. \eqref{series}, we get the series solution to Eq.
\eqref{homotopyeq1}
\begin{equation}\label{homotopyeqseriessol1}
u=-\frac{t}{2(t+C_2)}+\sum_{k=0}^\infty\frac{P_k(\xi)q^k}{(t+C_2)(t+C_2-\theta
tq)^k},
\end{equation}
and when further setting $q=1$, we have
\begin{equation}\label{boussisol1}
u=-\frac{t}{2(t+C_2)}+\sum_{k=0}^\infty\frac{P_k(\xi)}{(t+C_2)(t+C_2-\theta
t)^k},
\end{equation}
which is a homotopy series solution to the six-order boussinesq
equation.

The determination of similarity reduction equations depends on
finite equations in Eqs. \eqref{deformationeq} and
\eqref{similaritysol1}. It should be emphasized that all the previous
similarity reduction equations should be considered to remove the
terms $P_{0,\xi\xi\xi\xi}$, $P_{1,\xi\xi\xi\xi}$, $\cdots$,
$P_{k-1,\xi\xi\xi\xi}$ rather than $P_{0,\xi\xi\xi\xi\xi\xi}$,
$P_{1,\xi\xi\xi\xi\xi\xi}$, $\cdots$, $P_{k-1,\xi\xi\xi\xi\xi\xi}$
when we eliminate $u_k$ in the $k$th order approximate equation
\eqref{deformationeq} in terms of the similarity solutions
\eqref{similaritysol1}. We sum up the general formula for the
similarity reduction equations
\begin{eqnarray}\label{similarityeq1}
&&P_{k,\xi\xi\xi\xi}+2\sum_{i=0}^k(P_{k-i}P_{i,\xi\xi}+P_{k-i,\xi}P_{i,\xi})+(\frac{\xi^2}{4}+C_2)P_{k,\xi\xi}+\frac{1}{4}
(4k+7)\xi P_{k,\xi}\nonumber\\
&&-(k-1)C_2\theta\xi P_{k-1,\xi}+(k+1)(k+2)P_k-2(k+1)(k-1)C_2\theta P_{k-1}+(k-1)(k-2)(C_2\theta)^2P_{k-2}\nonumber\\
&&+C_2\delta_{k,0}+\sum_{i=0}^{k-1}\mu(\theta-1)(C_2\theta)^{k-1-i}
P_{i,\xi\xi\xi\xi\xi\xi}=0,\ (k=0,\ 1,\ \cdots)
\end{eqnarray}
with $P_{-2}=P_{-1}=0$.

When $k=0$, Eq.
\eqref{similarityeq1} is equivalent to the Painlev\'e IV type
equation. When $k>0$, specific forms of Eq. \eqref{similarityeq1}
depend on the solutions $P_0$, $P_1$, $\cdots$, $P_{k-1}$ and we can
rearrange the terms in Eq. \eqref{similarityeq1} as
$$
P_{k,\xi\xi\xi\xi}+2(P_kP_{0,\xi\xi}+P_0P_{k,\xi\xi})+4P_{k,\xi}P_{0,\xi}+(\frac{\xi^2}{4}+C_2)P_{k,\xi\xi}
$$
$$
+\frac{1}{4} (4k+7)\xi P_{k,\xi}+(k+1)(k+2)P_k=f_k(\xi),\ (k=1,\ 2,\
\cdots)\eqno(38')
$$
where $f_k(\xi)$ is a function of $\{P_0,\ P_1,\ \cdots,\ P_{k-1}\}$
\begin{displaymath}
f_k(\xi)=(k-1)C_2\theta\xi P_{k-1,\xi}+2(k+1)(k-1)C_2\theta
P_{k-1}-2\sum_{i=1}^{k-1}(P_{k-i}P_{i,\xi\xi}+P_{k-i,\xi}P_{i,\xi})
\end{displaymath}
\begin{displaymath}
-(k-1)(k-2)(C_2\theta)^2P_{k-2}-\sum_{i=0}^{k-1}\mu(\theta-1)(C_2\theta)^{k-1-i}
P_{i,\xi\xi\xi\xi\xi\xi}.
\end{displaymath}
Eq. $(38')$ is actually a fourth order linear variable coefficients ordinary differential
equation of $P_k$ when $P_0$, $P_1$, $\cdots$, $P_{k-1}$ are known.

\noindent\textit{Remark:} Taking $\theta=0$ and making the
transformations $P_k(\xi)=\mu^kQ_k(\xi)\ (k=0,\ 1,\ \cdots)$, we
change the similarity reduction equations \eqref{similarityeq1} and
the homotopy series solution \eqref{boussisol1} to
\begin{eqnarray}
&&Q_{k,\xi\xi\xi\xi}+2\sum_{i=0}^k(Q_{k-i}Q_{i,\xi\xi}+Q_{k-i,\xi}Q_{i,\xi})+(\frac{\xi^2}{4}+C_2)Q_{k,\xi\xi}+\frac{1}{4}
(4k+7)\xi Q_{k,\xi}\nonumber\\
&&+(k+1)(k+2)Q_k+C_2\delta_{k,0}-Q_{k-1,\xi\xi\xi\xi\xi\xi}=0,\ (k=0,\
1,\ \cdots)\label{apprsymsimilarityeq1}
\end{eqnarray}
and
\begin{equation}\label{apprsymseriessol1}
u=-\frac{t}{2(t+C_2)}+\sum_{k=0}^\infty\frac{\mu^kQ_k(\xi)}{(t+C_2)^{k+1}},
\end{equation}
respectively. It is easily seen that Eqs. \eqref{bouss3},
\eqref{apprsymsimilarityeq1} and \eqref{apprsymseriessol1} coincide
with Eqs. (4), (21) and (20) in \cite{Jiao} respectively.
\subsection{Homotopy symmetry reduction of the traveling wave form}
When $C_1=0$, from the characteristic equations \eqref{charaeq}, we get the similarity solutions
\begin{equation}\label{similaritysol2}
u_k=P_k(\xi),\ (k=0,\ 1,\ \cdots)
\end{equation}
where $\xi=t-\frac{C_2}{C_3}x$. We take an equivalent travelling
wave form $\xi=x+ct$ for the similarity variable with $c$ an
arbitrary velocity constant.

From Eqs. \eqref{series} and \eqref{series1}, we obtain a series
solution to the homotopy model \eqref{homotopyeq1}
\begin{equation}\label{homotopyeqseriessol2}
u=\sum_{k=0}^\infty q^kP_k(\xi),
\end{equation}
and a homotopy series solution to the six-order boussinesq equation
\begin{equation}\label{boussisol2}
u=\sum_{k=0}^\infty P_k(\xi).
\end{equation}

By substituting the similarity solutions \eqref{similaritysol2} to
approximate equations \eqref{deformationeq}, we get the similarity
reduction equations
\begin{equation}\label{similarityeq2}
P_{k,\xi\xi\xi\xi}+(c^2+1)P_{k,\xi\xi}+2\sum_{i=0}^k(P_{k-i}P_{i,\xi\xi}+P_{k-i,\xi}P_{i,\xi})+
\mu(\theta-1)\sum_{i=0}^{k-1}\theta^{k-1-i}P_{i,\xi\xi\xi\xi\xi\xi}=0,\
(k=0,\ 1,\ \cdots)
\end{equation}
which are equivalent to
$$
P_{k,\xi\xi}+(c^2+1)P_k+2(2-\delta_{k,0})P_kP_0=g_k(\xi),\ (k=0,\
1,\ \cdots)\eqno(44')
$$
where
\begin{displaymath}
g_k(\xi)=-2\sum_{i=1}^{k-1}P_{k-i}P_i-\mu(\theta-1)\sum_{i=0}^{k-1}\theta^{k-1-i}P_{i,\xi\xi\xi\xi}-a_k\xi-b_k,
\end{displaymath}
with $a_k$ and $b_k$ arbitrary integral constants. Eq. $(44')$ is a
second order linear variable coefficients ordinary differential equation of $P_k$.

\noindent\textit{Remark:} Taking $\theta=0$ and making the
transformation $P_k(\xi)=\mu^kQ_k(\xi)\ (k=0,\ 1,\ \cdots)$, we
reduce the similarity reduction equations \eqref{similarityeq2} and
the homotopy series solution \eqref{boussisol2} to
\begin{equation}\label{apprsymsimilarityeq2}
Q_{k,\xi\xi\xi\xi}+(c^2+1)Q_{k,\xi\xi}+2\sum_{i=0}^k(Q_{k-i}Q_{i,\xi\xi}+Q_{k-i,\xi}Q_{i,\xi})-
Q_{k-1,\xi\xi\xi\xi\xi\xi}=0,\ (k=0,\ 1,\ \cdots)
\end{equation}
and
\begin{equation}\label{apprsymseriessol2}
u=\sum_{k=0}^\infty \mu^kQ_k(\xi),
\end{equation}
respectively, with $Q_{-1}=0$. It is easily seen that Eqs.
\eqref{bouss3} and \eqref{apprsymseriessol2} are identical to Eqs.
(4) and (24) in \cite{Jiao} respectively. Integrating Eq.
\eqref{apprsymsimilarityeq2} twice, we obtain Eq. (25) in
\cite{Jiao}.

When $k=0$, Eq. \eqref{similarityeq2} is equivalent to the
Weierstrass elliptic equation and has the following solutions
\begin{eqnarray}
&&P_0(\xi)=4a^2-\frac{1}{2}(c^2+1)-6a^2\tanh^2(a\xi+b),\label{zerothtanhsol}\\
&&P_0(\xi)=2a^2(1-2m_1^2)-\frac{1}{2}(1+c^2)+6a^2m_1^2\mathrm{cn}^2(a\xi+b),\\
&&P_0(\xi)=-\frac{1}{2}(1+c^2)-6a^2\wp(a\xi+b,m_2,m_3),
\end{eqnarray}
where $a$, $b$, $m_1$, $m_2$ and $m_3$ are arbitrary constants.

When $k>0$ and $a_0=0$, Eq. $(44')$ can be integrated out one after
another
\begin{equation}
P_k=P_{0,\xi}\left[e_k+\int P_{0,\xi}^{-2}\left(c_k+\int
P_{0,\xi}g_k\mathrm{d}\xi\right)\mathrm{d}\xi\right],
\end{equation}
where $c_k$ and $e_k$ are arbitrary integral constants.

In the rest of this section, we discuss the solutions of similarity
reduction equations \eqref{similarityeq2} from hyperbolic tangent
function solution \eqref{zerothtanhsol}. We suppose that Eq.
\eqref{similarityeq2} have the hyperbolic tangent function solutions
\begin{equation}\label{kthtanhsol}
P_k(\xi)=\sum_{i=0}^{n_k}d_{k,i}\tanh^i\xi,\ (k=0,\ 1,\ \cdots)
\end{equation}
where all $d_{k,i}$ are constants to be determined. By balancing the
highest powers of $\tanh\xi$ from $P_{k-1,\xi\xi\xi\xi\xi\xi}$ and
$P_{k,\xi\xi}$ in Eq. \eqref{similarityeq2}, we have
$n_{k-1}+6=n_k+4$, with $n_0=2$ in Eq. \eqref{zerothtanhsol},
leading to $n_k=2(k+1)$.

The $k$th similarity reduction equation in Eq. \eqref{similarityeq2}
contains $P_0$, $P_1$, $\cdots$, $P_k$, so that $P_0$, $P_1$,
$\cdots$, $P_k$ can be solved one after another starting from Eq.
\eqref{zerothtanhsol}. For the $k$th equation in Eq.
\eqref{similarityeq2}, $P_0$, $P_1$, $\cdots$, $P_{k-1}$ are known
and inserted into Eq. \eqref{similarityeq2} together with Eq.
\eqref{kthtanhsol}. Matching the coefficients of different powers of
$\tanh\xi$, we get a system of algebraic equations with respect to
$d_{k,0}$, $d_{k,1}$, $\cdots$, $d_{k,2(k+1)}$ from which we can
construct Eq. \eqref{kthtanhsol}.

We list $d_{k,i}$ in Eq. \eqref{kthtanhsol} up to $k=3$

\noindent$k=1:$
\begin{displaymath}
d_{1,0}=22\mu a^4(\theta-1),\ d_{1,1}=p_1,\ d_{1,2}=-120\mu
a^4(\theta-1),\ d_{1,3}=-p_1,\ d_{1,4}=90\mu a^4(\theta-1),
\end{displaymath}
$k=2:$
\begin{displaymath}
d_{2,0}=180\mu^2a^6(\theta-1)^2+22\mu
a^4\theta(\theta-1)-\frac{p_1^2}{24a^2}, d_{2,3}=-30\mu
a^2p_1(\theta-1)-p_2,
\end{displaymath}
\begin{displaymath}
d_{2,1}=p_2,\ d_{2,2}=-2970\mu^2a^6(\theta-1)^2-120\mu
a^4\theta(\theta-1)+\frac{p_1^2}{6a^2},\ d_{2,5}=30\mu
a^2p_1(\theta-1),
\end{displaymath}
\begin{displaymath}
d_{2,4}=5580\mu^2a^6(\theta-1)^2+90\mu
a^4\theta(\theta-1)-\frac{p_1^2}{8a^2},\
d_{2,6}=-2790\mu^2a^6(\theta-1)^2,
\end{displaymath}
$k=3:$
\begin{displaymath}
d_{3,0}=6030\mu^3a^8(\theta-1)^3+360\mu^2a^6\theta(\theta-1)^2+22\mu
a^4\theta^2(\theta-1)+\frac{5}{6}\mu
p_1^2(\theta-1)-\frac{p_1p_2}{12a^2},
\end{displaymath}
\begin{displaymath}
d_{3,2}=-118800\mu^3a^8(\theta-1)^3-5940\mu^2a^6\theta(\theta-1)^2-120\mu
a^4\theta^2(\theta-1)+\frac{5}{12}\mu
p_1^2(\theta-1)+\frac{p_1p_2}{3a^2},
\end{displaymath}
\begin{displaymath}
d_{3,3}=-1260\mu^2a^4p_1(\theta-1)^2-30\mu
a^2(p_1\theta+p_2)(\theta-1)-p_3+\frac{p_1^3}{72a^4},
\end{displaymath}
\begin{displaymath}
d_{3,4}=361080\mu^3a^8(\theta-1)^3+11160\mu^2a^6\theta(\theta-1)^2+90\mu
a^4\theta^2(\theta-1)-\frac{15}{2}\mu
p_1^2(\theta-1)-\frac{p_1p_2}{4a^2},
\end{displaymath}
\begin{displaymath}
d_{3,5}=2655\mu^2a^4p_1(\theta-1)^2+30\mu
a^2(p_1\theta+p_2)(\theta-1)-\frac{p_1^3}{72a^4},
\end{displaymath}
\begin{displaymath}
d_{3,6}=-397296\mu^3a^8(\theta-1)^3-5580\mu^2a^6\theta(\theta-1)^2+\frac{25}{4}\mu
p_1^2(\theta-1),
\end{displaymath}
\begin{displaymath}
d_{3,1}=p_3,\ d_{3,7}=-1395\mu^2a^4p_1(\theta-1)^2,\
d_{3,8}=148986\mu^3a^8(\theta-1)^3,
\end{displaymath}
where $p_1$, $p_2$ and $p_3$ are arbitrary constants.

We specify the parameters by $a=1$, $b=0$, $c=1$,
$\mu=0.1$, $p_1=0$, $p_2=0$, $p_3=0$, $p_4=0$ and $p_5=0$.
To see the relationship between the convergence of the homotopy
series solutions and the parameter $\theta$, we choose four values
$0.4$, $0.8$, $1.2$ and $1.6$ for $\theta$ and display the plots of
$|P_k(\xi)|\ (k=1,\ 2,\ 3)$ for $-5<x<3$ and $t=1$ in Figure 1 where
the dotted line, the dashed line and the solid line represent
$|P_1(\xi)|$, $|P_2(\xi)|$ and $|P_3(\xi)|$ respectively.
\begin{figure}
\centering \subfigure[$\theta$=0.4]{
\includegraphics[width=0.4\textwidth]{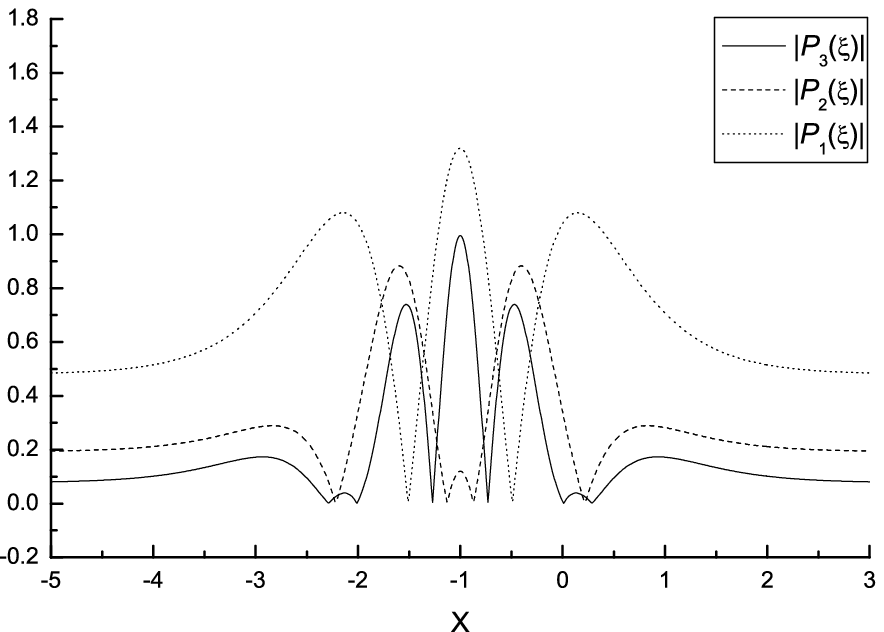}}
\subfigure[$\theta$=0.8]{
\includegraphics[width=0.4\textwidth]{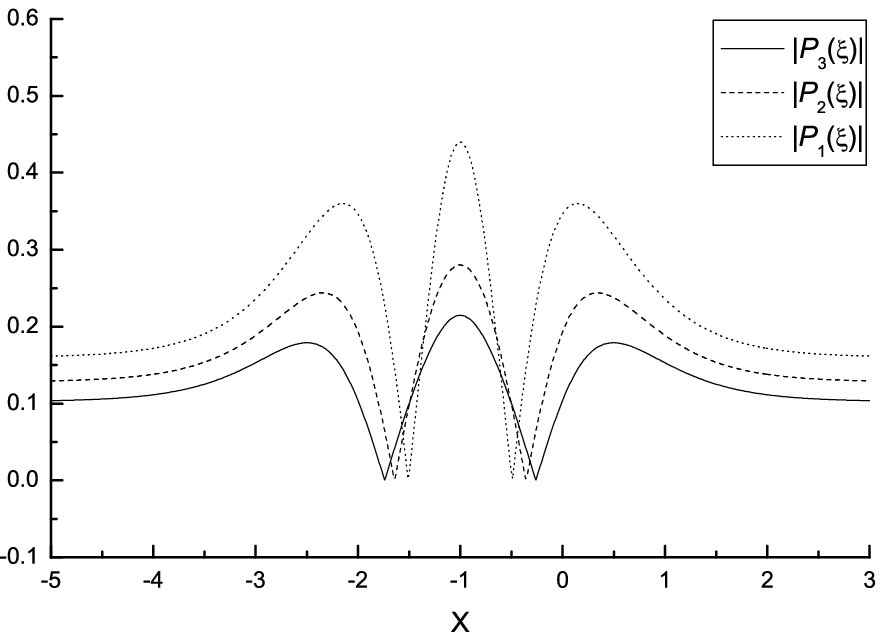}}\\
\centering \subfigure[$\theta$=1.2]{
\includegraphics[width=0.4\textwidth]{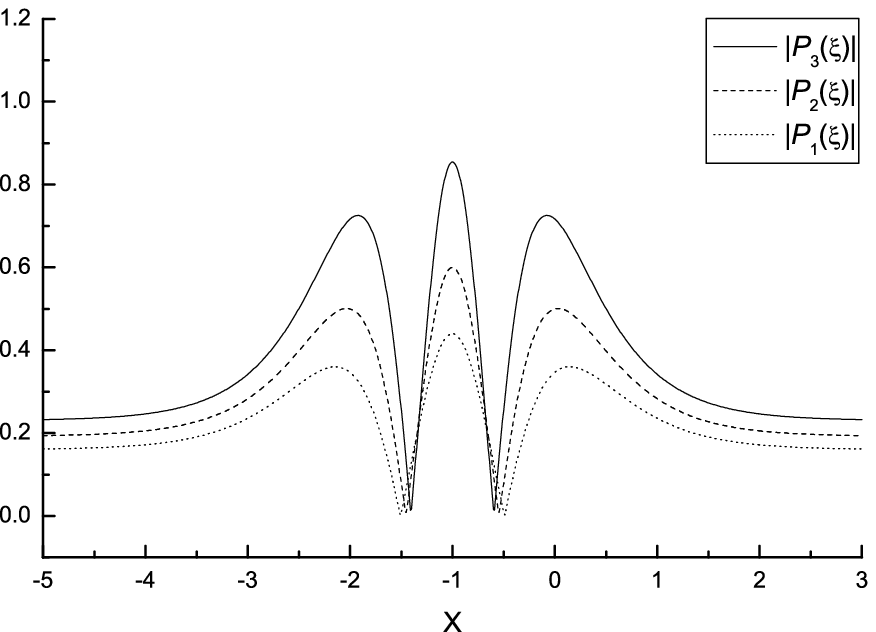}}
\subfigure[$\theta$=1.6]{
\includegraphics[width=0.4\textwidth]{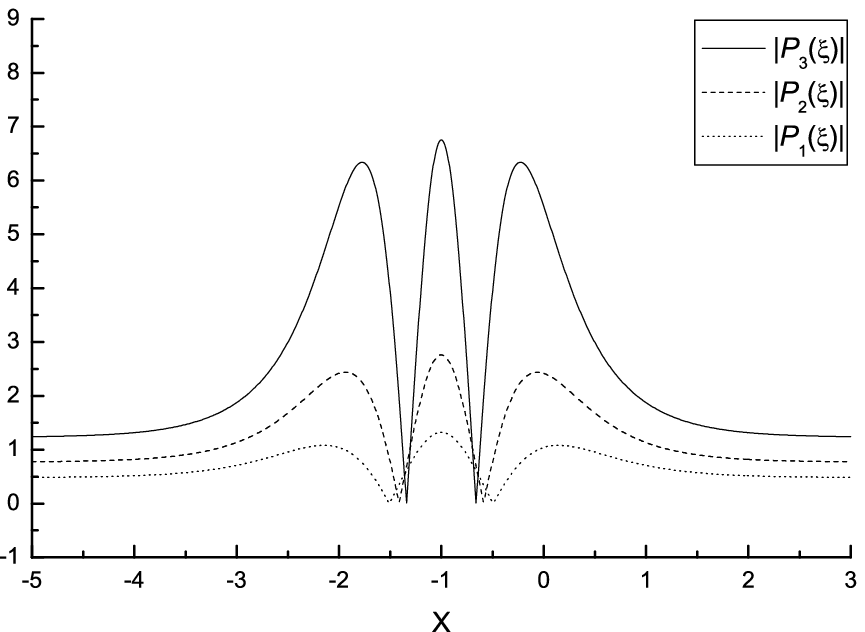}}
\caption{Plots of $|P_k(\xi)|$ in Eq. \eqref{kthtanhsol} for $k=1,\
2,\ 3$.}
\end{figure}
The necessary condition for the convergence of homotopy series solutions
\eqref{boussisol2} is $\lim_{k \to\infty}|P_k(\xi)|=0$ and the convergence regions correspond to lower
solid lines in Figure 1.

For (a) and (b) in Figure 1, the convergence region of the homotopy
series solutions \eqref{boussisol2} for $\theta=0.8$ is wider than
the convergence region for $\theta=0.4$. The homotopy series
solutions \eqref{boussisol2} corresponding to (c) and (d) in Figure
1 are divergent, with the homotopy series solutions for $\theta=1.6$
diverging faster than the homotopy series solutions for
$\theta=1.2$.

\noindent\textit{Remark:} When $\theta<1\ (\lambda>0)$, the
convergence regions of the homotopy series solutions
\eqref{boussisol2} grow wider provided that $\theta \to 1\ (\lambda
\to 0)$. When $\theta>1\ (\lambda<0)$, the homotopy series solutions
\eqref{boussisol2} are divergent and the larger the values of
$\theta$ (the smaller the values of $\lambda$), the faster the
homotopy series solutions diverge.
\section{Summary and discussion}
In the framework of approximate homotopy symmetry method, we
investigated the six-order boussinesq equation and summarized the
similarity reduction solutions and similarity reduction equations
for approximate equations of different orders. The homotopy series
solutions to the six-order boussinesq equation were derived.

Zero-order similarity reduction equations are equivalent to
Painlev\'e IV type equation and Weierstrass elliptic equation.
$k$-order similarity reduction equations are linear variable coefficients ordinary
differential equations of $P_k(\xi)$ which depend on particular
solutions of the previous similarity reduction equations from
zero-order to $(k-1)$-order.

For homotopy symmetry reduction of the traveling wave form, we
constructed hyperbolic tangent function solutions to $k$-order
similarity reduction equations $(k=1,\ 2,\ 3)$.
$|P_k(\xi)|$ were plotted for different values of $\lambda$.

The auxiliary parameter $\lambda$ dominates the convergence regions
of the homotopy series solutions. For $\lambda>0$, the convergence
regions get wider when $\lambda \to 0$. For $\lambda<0$, the
homotopy series solutions are divergent and smaller values of
$\lambda$ correspond to faster diverging homotopy series solutions.
When $\lambda \to 0$, Eq. \eqref{homotopyeq1} tends to the
boussinesq equation. Accordingly, the homotopy series solutions to
the six-order boussinesq equation behave more and more like the
solutions to the boussinesq equation. This may account for the role
of $\lambda$ in controlling the convergence regions of homotopy
series solutions.

The approximate symmetry method is valid only if the nonlinear
problems contain small parameters. In contrast, the auxiliary
parameter $\lambda$ for the approximate homotopy symmetry method is
artificial introduced. In other words, the approximate homotopy
symmetry method is applicable to those nonlinear problems that
contain no parameters at all. For the six-order boussinesq equation,
the convergence of all approximate symmetry series solutions in Ref.
\cite{Jiao} is influenced by the parameter $\mu$. However, the
parameter $\mu$ is not explicitly included in the homotopy series
solutions \eqref{boussisol1} and \eqref{boussisol2}, showing that
the convergence of the homotopy series solutions is not directly
affected by large parameters.

Moreover, the series solutions from approximate symmetry method can
also be retrieved by approximate homotopy symmetry method.
Therefore, the approximate homotopy symmetry method is superior to
the approximate symmetry method and provides an effective analytical
tool in constructing series solutions for many nonlinear problems.
\newline

\noindent{\bf Acknowledgement}

The authors are grateful to Prof. Shijun Liao for helpful discussions in finishing the manuscript.
The work was supported by the National Natural Science Foundations
of China (Nos. 10735030, 10475055, 10675065 and 90503006), National
Basic Research Program of China (973 Program 2007CB814800) and
PCSIRT (IRT0734), the Research Fund of Postdoctoral of China (No.
20070410727) and Specialized Research Fund for the Doctoral Program
of Higher Education (No. 20070248120).

\end{document}